\documentclass[10pt]{article}
\usepackage{epsf} 
\usepackage{amsmath}
\usepackage{amssymb}
\usepackage{epsfig}
\usepackage{latexsym}
\usepackage{amsfonts}
\usepackage{graphicx}%
\usepackage{varioref}
\usepackage{ifthen}
\begin{document}
\parindent 0mm 
\setlength{\parskip}{\baselineskip} 
\pagenumbering{arabic} 
\setcounter{page}{1}
\mbox{ }
\rightline{UCT-TP-286/2010}
\newline
\newline
\rightline{March 2011}
\newline

\begin{center}
\Large \textbf{Quark masses in QCD: a progress report}
{\footnote{{\LARGE {\footnotesize Invited review paper to be published in Modern Physics Letters A. }}}}
\end{center}
\begin{center}
C. A. Dominguez
\end{center}

\begin{center}
Centre for Theoretical and Mathematical Physics,  
University of Cape Town, Rondebosch 7700, South Africa, and Department of Physics, Stellenbosch University, Stellenbosch 7600, South Africa
\end{center}

\begin{center}
\textbf{Abstract}
\end{center}

\noindent
Recent progress on QCD sum rule determinations of the light and  heavy quark masses is reported. In the light quark sector a major breakthrough has been made recently in connection with the historical systematic uncertainties due to a lack of  experimental information on the pseudoscalar  resonance spectral functions. It is now possible to suppress this contribution to the $1 \%$ level  by using suitable integration kernels in Finite Energy QCD sum rules. This allows to determine the up-, down-, and strange-quark masses with an unprecedented precision of some $8 - 10\, \%$. Further reduction of this uncertainty will be possible with improved accuracy in the strong coupling, now the main source of error.
In the heavy quark sector, the availability of experimental data in the vector channel, and the use of suitable multipurpose integration kernels allows to increase the accuracy of the charm- and bottom-quarks masses to the $1 \%$ level.

\section{Introduction}
Quark and gluon confinement in Quantum Chromodynamics (QCD) precludes direct experimental measurements of the fundamental QCD parameters, i.e. the strong interaction coupling and the quark masses. Hence, in order to determine these parameters analytically one needs to relate them to experimentally measurable quantities. Alternatively,  simulations of QCD on a lattice provide increasingly accurate numerical values for these parameters, but little if any insight into their origin. The first approach relies on the intimate relation between QCD Green functions, in particular their Operator Product Expansion (OPE) beyond perturbation theory, and their hadronic counterparts. This relation follows from Cauchy's theorem in the complex energy plane, and is  known as the QCD sum rule technique \cite{QCDSR}. In addition to producing numerical values for the QCD parameters, this method provides a detailed breakdown of the relative impact of the various dynamical contributions. For instance, the strong coupling at the scale of the $\tau$-lepton mass  essentially follows from the relation between the experimentally measured $\tau$ ratio, $R_\tau$, and a contour integral involving the perturbative QCD (PQCD) expression of the $V+A$ correlator. This is the cleanest, most transparent, and model independent determination of the strong coupling \cite{PICH1}-\cite{PICH2}. It also allows to gauge the impact of each individual term in PQCD, up to the currently known five-loop order.
Similarly, in the case of the quark masses one considers a QCD correlation function which on the one hand involves the quark masses and other QCD parameters, and on the other hand it involves a measurable (hadronic) spectral function. Using Cauchy's theorem to relate both representations, the quark masses  become a function of QCD parameters, e.g. the strong coupling, some vacuum condensates reflecting confinement, etc., and measurable hadronic parameters. The virtue of this approach is that it provides a breakdown of each contribution to the final value of the quark masses. More importantly, it allows to tune the relative weight of each of these contributions by introducing suitable integration kernels. This last feature has been used recently to solve the historical problem of the systematic uncertainties affecting light quark mass determinations, to be discussed in this report.\\
In the case of the light quark masses  the ideal Green function is the pseudoscalar current correlator. This contains the square of the quark masses as an overall factor multiplying the PQCD expansion, and the leading power corrections in the OPE. Unfortunately, this correlator is not realistically accessible experimentally beyond the pseudoscalar meson pole. While the existence of at least two radial excitations
of the pion and the kaon are known from hadronic interaction data, this information is hardly enough to reconstruct the full  spectral functions. In spite of many attempts over the years to model them, there remains an  unknown systematic uncertainty that has plagued light quark mass determinations from QCD sum rules. The use of the vector current correlator, for which there is plenty of experimental data from $\tau$ decays and $e^+ e^-$ annihilation, is not a realistic option for the light quarks as their masses enter as sub-leading terms in the OPE. The scalar correlator, involving the square of quark mass differences, at some stage offered some promise  for determining the strange quark mass with reduced systematic uncertainties. This was due to the availability of data on $K-\pi$ phase shifts. Unfortunately, these data do not fully determine the hadronic spectral function. The latter can be reconstructed from phase shift data only after substantial theoretical manipulations, implying a large unknown systematic uncertainty. A breakthrough has been made recently  by introducing an integration kernel in the contour integral in the complex energy plane. This allows to suppress substantially the unknown hadronic resonance contribution to the pseudoscalar current correlator. As it follows from Cauchy's theorem, this suppression implies that the quarks masses are determined essentially from the well known pseudoscalar meson pole and PQCD (well known up to five-loop level). In this way it has  been possible to reduce the hadronic resonance contribution to the 1\% level, allowing for an unprecedented accuracy of some $8 - 10\, \%$ in the values of the up-, down-, and strange-quark masses. Further improvement on this accuracy will be possible with further reduction of the uncertainty in the strong coupling, now the main source of error.\\
The determination of the charm- and bottom-quark masses has been free of  systematic uncertainties due to the hadronic resonance sector, as there is plenty of experimental information in the vector channel from $e^+ e^-$ annihilation into hadrons. One problem, though, is that the massive vector current correlator is not known in PQCD to the same level as the light pseudoscalar correlation function. Nevertheless, substantial theoretical progress has been made over the years leading to extremely accurate charm- and bottom-quark masses. The novel idea of introducing suitable integration kernels in Cauchy's contour integrals, as described above, has also been used recently as a way of improving accuracy in the heavy-quark sector. For instance, kernels can be used to suppress regions where the data is either not as accurate, or simply unavailable. This will also be reported here.\\
The paper is organized as follows. First, determinations of quark-mass ratios from various hadronic data, as well as from chiral perturbation theory, will be reviewed in Section 2. These ratios are quite useful as consistency checks for results from QCD sum rules. Section 3 describes the OPE beyond perturbation theory, one of the two pillars of QCD sum rules. Section 4 discusses quark-hadron duality and finite energy sum rules.
These sum rules weighted by suitable integration kernels will be analyzed in the light quark sector in Section 5. In particular it will be shown how this technique unveils the subjacent hadronic systematic uncertainty plaguing  light quark mass determinations for the past thirty years. In Section 6 recent progress on charm- and bottom-quark mass determinations will be reported. Comparison with lattice QCD results for all quark masses will also be made. Finally, Section 7  provides  a very short summary of this report.\\
As an important disclaimer, this paper is not a comprehensive review of past quark mass determinations from QCD sum rules. It is, rather, a report on recent progress on the subject. Given that past determinations of light quark masses were affected by unknown systematic uncertainties, essentially from the hadronic resonance sector, it makes little or no sense to review them once the main uncertainty has been exposed. Any agreement between values affected by this  uncertainty and current results, free of it, would only be fortuitous. In fact, once this hadronic resonance uncertainty is removed, the values of all three light quark masses get reduced by some 15 - 20 \%, a clear sign of a systematic uncertainty acting in only one direction. Last but not least,  light quark masses from QCD sum rules before 2006 employed correlators  up to at most four-loop level in PQCD, together with superseded values of the strong coupling.
\section{Quark mass ratios}
Quark masses actually precede QCD by a number of years, albeit under the guise of {\it current algebra quark masses}, which clearly lacked today's detailed understanding of quark-mass renormalization. In fact, the study of global ${\mbox{SU}}(3) \times {\mbox{SU}}(3)$ chiral symmetry realized \'{a} la Nambu-Goldstone, and its breaking down to ${\mbox{SU}}(2) \times {\mbox{SU}}(2)$, followed by a breaking down to ${\mbox{SU}}(2)$, and finally to ${\mbox{U}}(1)$  was first done using the strong interaction Hamiltonian \cite{CA1}-\cite{HLR}
\begin{equation}
H(x)\, =\, H_0(x)\, + \,\epsilon_0\; u_0(x)\, + \,\epsilon_3\; u_3(x)\, + \,\epsilon_8\; u_8(x) \;.
\end{equation}
The term $H_0(x)$ above is $\mbox{SU}(3) \times \mbox{SU}(3)$ invariant, the $\epsilon_{0,3,8}$  are symmetry breaking parameters, and the scalar densities $u_{0,3,8}(x)$  transform according to the $3 \,\overline{3} \;+\;\overline{3} \,3$ representation of $\mbox{SU}(3) \times \mbox{SU}(3)$. In modern language, $\epsilon_8$ is related to the strange quark mass $m_s$, and $\epsilon_3$ to the difference between the down- and the up-quark masses $m_d-m_u$, while the scalar densities are related to products of quark-anti-quark field operators.  For instance, the ratio of $\mbox{SU}(3)$ breaking to $\mbox{SU}(2)$ breaking is given by
\begin{equation}
R \equiv \frac{m_s - m_{ud}}{m_d - m_u} = \frac{\sqrt{3}}{2} \; \frac{\epsilon_8}{\epsilon_3} \;,
\end{equation}
where $m_{ud} \equiv (m_u + m_d)/2$. In the pre-QCD era many relations for quark-mass ratios were obtained from hadron mass ratios, as well as from other hadronic information, e.g. $\eta \rightarrow 3 \pi$, $K_{l 3}$ decay, etc. \cite{HLR}.  To mention a pioneering  determination of the ratio $R$ above, from a solution to the $\eta \rightarrow 3 \pi$ puzzle proposed in  \cite{CAD1}  it followed \cite{CAD2}  $R^{-1} = 0.020 \pm 0.002$, in remarkable agreement with a later determination based on baryon mass splittings \cite{MZ} $R^{-1} = 0.021 \pm 0.003$. With the advent of chiral perturbation theory (CHPT) \cite{HP}-\cite{HLR}, \cite{CHPT1}-\cite{CHPT2}, certain quark mass ratios turned out to be renormalization scale independent to leading order, and could be expressed in terms of pseudoscalar meson mass ratios \cite{HLR},\cite{CHPT1}-\cite{SW}, e.g.
\begin{equation}
\frac{m_u}{m_d} = \frac{M_{K^+}^2 - M_{K^0}^2 + 2 M_{\pi^0}^2 - M_{\pi^+}^2}{M_{K^0}^2 - M_{K^+}^2 + M_{\pi^+}^2} = 0.56
\end{equation}
\begin{equation}
\frac{m_s}{m_d} = \frac{M_{K^+}^2 + M_{K^0}^2 - M_{\pi^+}^2}{M_{K^0}^2 - M_{K^+}^2 + M_{\pi^+}^2} = 20.2 \;,
\end{equation}
where the numerical results follow after some subtle corrections due to electromagnetic self energies \cite{CHPT2}.
Beyond leading order in CHPT things become complicated. At next to leading order (NLO) the only  parameter free relation is
\begin{equation}
Q^2 \equiv \frac{m_s^2 - m_{ud}^2}{m_d^2 - m_u^2} =
\frac{M_{K}^2 - M_{\pi}^2}{M_{K^0}^2 - M_{K^+}^2}\; \frac{M_K^2}{M_{\pi}^2} \;. 
\end{equation}
Other quark mass ratios  at NLO and beyond depend on the renormalization scale, as well as on some CHPT low energy constants which need to be determined independently \cite{CHPT1}-\cite{CHPT2}. After taking into account electromagnetic self energies, Eq.(5) gives \cite{CHPT1}  $Q = 24.3$, while a recent analysis of $\eta \rightarrow 3 \pi$ \cite{CHPT1}, \cite{GC} gives
\begin{equation}
Q = 22.3 \pm 0.8 \;.
\end{equation}
The ratios $R$, Eq.(2), and $Q$, Eq.(5), together with the leading order ratios Eqs.(3)-(4), will prove useful for comparisons with QCD sum rule results. An additional useful quark mass ratio involving the
ratios Eqs.(3)-(4) is
\begin{equation}
r_s \equiv \frac{m_s}{m_{ud}} = \frac{2\; m_s/m_d}{1 + m_u/m_d} = 28.1 \pm 1.3 \;,
\end{equation}
where the numerical value follows from the NLO CHPT relation \cite{CHPT1}, to be compared with the LO result from Eqs.(3)-(4), $r_s = 25.9$, and a large $N_c$ estimate \cite{eta} $r_s = 26.6 \pm 1.6$.
\section{Operator product expansion beyond perturbation theory}
The OPE beyond perturbation theory in QCD, one of the two pillars of the sum rule technique, is an effective
tool to introduce quark-gluon confinement dynamics. It is not a model, but rather a parametrization of quark and gluon propagator corrections due to confinement, done in a rigorous renormalizable quantum field theory framework. Let us consider a typical object in QCD in the form of the two-point function, or current correlator
\begin{equation}
\Pi(q^2)\,=\,i\; \int \,d^4 x \; e^{iqx} \; <0|\,T(J(x)\,J(0))\,|0 >,
\end{equation}
where the local current $J(x)$ is built from the quark and gluon fields entering the QCD Lagrangian. Equivalently, this current can also be written in terms of hadronic fields with the same quantum numbers. A relation between the two representations follows from Cauchy's theorem in the complex energy (squared) plane. This is often referred to as quark-hadron duality, the second pillar of the QCD sum rules method to be discussed in the next section.
The QCD correlator, Eq.(8),  contains a perturbative piece (PQCD), and a non perturbative one mostly reflecting quark-gluon confinement. The leading order in PQCD is shown in Fig.1.  Since confinement has not been proven analytically in QCD, its effects  can only be introduced effectively, e.g. by parameterizing quark and gluon propagator corrections in terms of vacuum condensates. This is done as follows. In the case of the quark propagator
\begin{equation}
S_F (p) = \frac{i}{\not{p} - m}\;\;\Longrightarrow \;\;\frac{i}{\not{p} - m + \Sigma(p^2)} \;, 
\end{equation}
the  propagator correction $\Sigma(p^2)$  contains the information on confinement, a purely non-perturbative effect. One expects this correction to peak at and near the quark mass-shell, e.g. for $p \simeq 0$ in the case of light quarks. Effectively, this can be viewed as in Fig. 2, where the (infrared) quarks in the loop have zero momentum and interact strongly with the physical QCD vacuum. This effect is then parameterized in terms of the quark condensate $\langle 0| \bar{q}(0) q(0) | 0 \rangle$.
\begin{figure}[ht]
\begin{center}
  \includegraphics[height=.1\textheight]{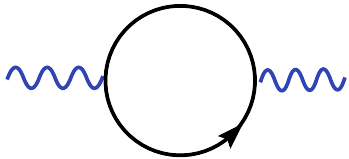}
  \caption{\footnotesize{
  Leading order PQCD correlator. All values of the four-momentum of the quarks in the loop are allowed. The blue wiggly line represents the current of momentum $q$ ($-q^2 >> 0$).}}
  \label{fig:figure1}
\end{center}
\end{figure}
Similarly, in the case of the gluon propagator 
\begin{equation}
D_F (k) = \frac{i}{k^2}\;\;\Longrightarrow \;\;\frac{i}{k^2 + \Lambda(k^2)} \;,
\end{equation}
the propagator correction $\Lambda(k^2)$ will peak at $k\simeq 0$, and the effect of confinement in this case can be parameterized by the gluon condensate $\langle 0| \alpha_s\; \vec{G}^{\mu\nu} \,\cdot\, \vec{G}_{\mu\nu}|0\rangle$ (see Fig.3).
In addition to the quark and the gluon condensate there is a plethora of higher order condensates entering the OPE of the current correlator at short distances, i.e.
\begin{equation}
\Pi(q^2)|_{QCD}\,=\, C_0\,\hat{I} \,+\,\sum_{N=0}\;C_{2N+2}(q^2,\mu^2)\;\langle0|\hat{O}_{2N+2}(\mu^2)|0\rangle \;,
\end{equation}
where $\mu^2$ is the renormalization scale, and where the Wilson coefficients in this expansion, 
$ C_{2N+2}(q^2,\mu^2)$,  depend on the Lorentz indices and quantum numbers of $J(x)$ and  of the local gauge invariant operators $\hat{O}_N$ built from the quark and gluon fields. These operators are ordered by increasing dimensionality and the Wilson coefficients, calculable in PQCD, fall off by corresponding powers of $-q^2$. In other words, this OPE achieves a factorization of short distance effects encapsulated in the Wilson coefficients, and long distance dynamics present in the vacuum condensates.
Since there are no gauge invariant operators of dimension $d=2$ involving the quark and gluon fields in QCD, it is normally assumed that the OPE starts at dimension $d=4$. This is supported by results from QCD sum rule analyses of $\tau$-lepton decay data, which show no evidence of $d=2$ operators \cite{C2a}-\cite{C2b}.
\begin{figure}[ht]
\begin{center}
\includegraphics[height=.06\textheight]{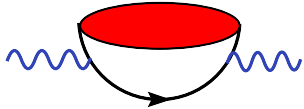}
\caption{\footnotesize{Quark propagator modification due to (infrared) quarks interacting with the physical QCD vacuum, and involving the quark condensate. Large momentum flows through the bottom propagator.}}
\label{fig:figure2}
\end{center}
\end{figure}
The unit operator $\hat{I}$ in Eq.(11) has dimension $d=0$ and $C_0 \hat{I}$ stands for the purely perturbative contribution. The Wilson coefficients as well as the vacuum condensates depend on the renormalization scale. For light quarks, and for the leading $d=4$ terms in Eq.(11), the $\mu^2$ dependence of the quark mass cancels the corresponding dependence of the quark condensate, so that this contribution is a renormalization group (RG) invariant. Similarly, the gluon condensate is also a RG invariant, hence once determined in some channel these condensates can be used throughout.
The numerical values of the vacuum condensates cannot be calculated analytically from first principles as this would be tantamount to solving QCD exactly.
One exception is that of the quark condensate which enters in the Gell-Mann-Oakes-Renner relation, a QCD low energy theorem following from the global chiral symmetry of the QCD Lagrangian \cite{GMOR}. Otherwise, it is possible to extract values for the leading vacuum condensates using QCD sum rules together with experimental data, e.g. $e^+ e^-$ annihilation into hadrons, and hadronic decays of the $\tau$-lepton. Alternatively, as lattice QCD  improves in accuracy it should become a valuable source of information on these condensates.\\
\begin{figure}[ht]
\begin{center}
\includegraphics[height=.12\textheight]{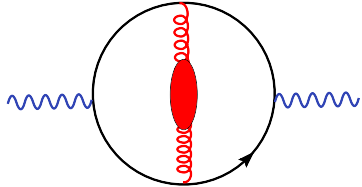}
\caption{\footnotesize{Gluon propagator modification due to (infrared) gluons interacting with the physical QCD vacuum, and involving the gluon condensate. Large momentum flows through the quark propagators.}}
\label{fig:figure3}
\end{center}
\end{figure}
\section{Quark-hadron duality and finite energy QCD sum rules}
Turning to the hadronic sector, bound states and resonances appear in the complex energy (squared) plane (s-plane) as poles on the real axis, and singularities in the second Riemann sheet, respectively. All these singularities lead to a discontinuity across the positive real  axis. Choosing an integration contour as shown in Fig. 4, and given that there are no other singularities in the complex s-plane, Cauchy's theorem leads to the finite energy sum rule (FESR)
\begin{figure}[ht]
\begin{center}
  \includegraphics[height=.25\textheight]{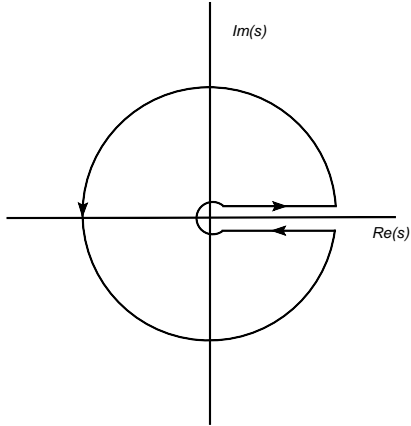}
  \caption{\footnotesize{Integration contour in the complex s-plane. The discontinuity across the real axis brings in the hadronic spectral function, while integration around the circle involves the QCD correlator. The radius of the circle is $s_0$, the onset of QCD.}}
\label{fig:figure4}
\end{center}
\end{figure}
\begin{equation}
\int_{\mathrm{sth}}^{s_0} ds\; \frac{1}{\pi}\; f(s) \;Im \,\Pi(s)|_{HAD} \; = \; -\, \frac{1}{2 \pi i} \; \oint_{C(|s_0|) }\, ds \;f(s) \;\Pi(s)|_{QCD} \;,
\end{equation}
where $f(s)$ is an arbitrary (analytic) function, $s_{th}$ is the hadronic threshold, and the finite radius of the circle, $s_0$, is large enough for QCD and the OPE to be used on the circle. 
Physical observables determined from FESR should be independent of $s_0$. In practice, though, 
this  is not exact, and there is usually a region of stability where
observables are fairly independent of $s_0$, typically somewhere inside the range $s_0 \simeq 1 - 4 \; \mbox{GeV}^2$. 
Equation (12) is the mathematical statement of what is usually referred to as quark-hadron duality. Since QCD is not valid in the time-like region ($s \geq 0$), in principle there is a possibility of problems on the circle near the real axis (duality violations), to be discussed shortly (this issue was identified very early in \cite{Shankar} long before the present formulation of QCD sum rules).
The right hand side  of this FESR involves the QCD correlator which is expressed in terms of the OPE as in Eq.(11). The left hand side involves the hadronic spectral function which is written as
\begin{equation} 
Im \,\Pi(s)|_{HAD}\,=\, Im \,\Pi(s)|_{POLE}\,+\, Im \,\Pi(s)|_{RES} \,\theta(s_0-s)\,+\, Im\, \Pi(s)|_{PQCD}\,\theta(s-s_0) \;,
\end{equation}
where the ground state pole is followed by the resonances which merge smoothly into the hadronic continuum above some threshold $s_0$. This continuum is expected to be well represented by PQCD if $s_0$ is large enough. Hence, if one were to consider an integration contour in Eq.(12) extending to infinity, the cancellation between the hadronic continuum on the left hand side and the PQCD contribution on the right hand side, would render the sum rule a FESR. 
\begin{figure}[ht]
\begin{center}
  \includegraphics[height=.28\textheight]{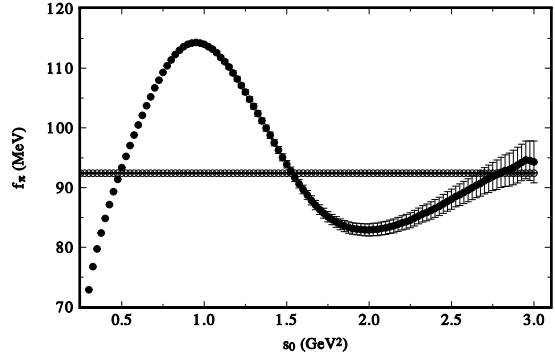}
  \caption{ \footnotesize{Results for $f_\pi$ from the standard  FESR in the axial-vector channel, Eq. (14) with N=0, with no dimension $d=2$ term, and using  CIPT, with $\Lambda_{QCD} = 365 \,\mbox{MeV}$ ($\alpha_s(M_\tau)= 0.335$). The straight line  is the experimental value of $f_\pi$, and the points are the integrated data with the experimental errors.}}
  \label{fig:figure5}
  \end{center}
\end{figure}
The integration in the complex s-plane of the QCD correlator is usually carried out in two different ways, Fixed Order Perturbation Theory (FOPT) and Contour Improved Perturbation Theory (CIPT). The first method treats running quark masses and the strong coupling as fixed at a given value of $s_0$. After integrating all logarithmic terms ($\ln(-s/\mu^2)$) the RG improvement is achieved by setting the renormalization scale to $\mu^2 = - s_0$. In CIPT the RG improvement is performed before integration, thus eliminating logarithmic terms, and the running quark masses and strong coupling are integrated  around the circle. This requires solving numerically the RGE for the quark masses and the coupling at each point on the circle.
The FESR Eq.(12) with $f(s)=1$ and  in FOPT can be written as
\begin{equation}
(-)^N \, C_{2N+2} \, \langle0| \hat{O}_{2N+2}|0 \rangle =  \int_0^{s_0} \,ds\, s^N \, \frac{1}{\pi}\, Im \,\Pi(s)|_{HAD} \,-\, s_0^{N+1} \; M_{2N+2}(s_0) \, ,
\end{equation}
where the dimensionless PQCD moments $M_{2N+2}(s_0)$ are given by
\begin{equation}
M_{2N+2}(s_0) = \frac{1}{s_0^{(N+1)}} \, \int_0^{s_0}\, ds\,s^N \, \frac{1}{\pi} \, Im \, \Pi(s)|_{PQCD}\;.
\end{equation}
\begin{figure}[ht]
\begin{center}
  \includegraphics[height=.35\textheight]{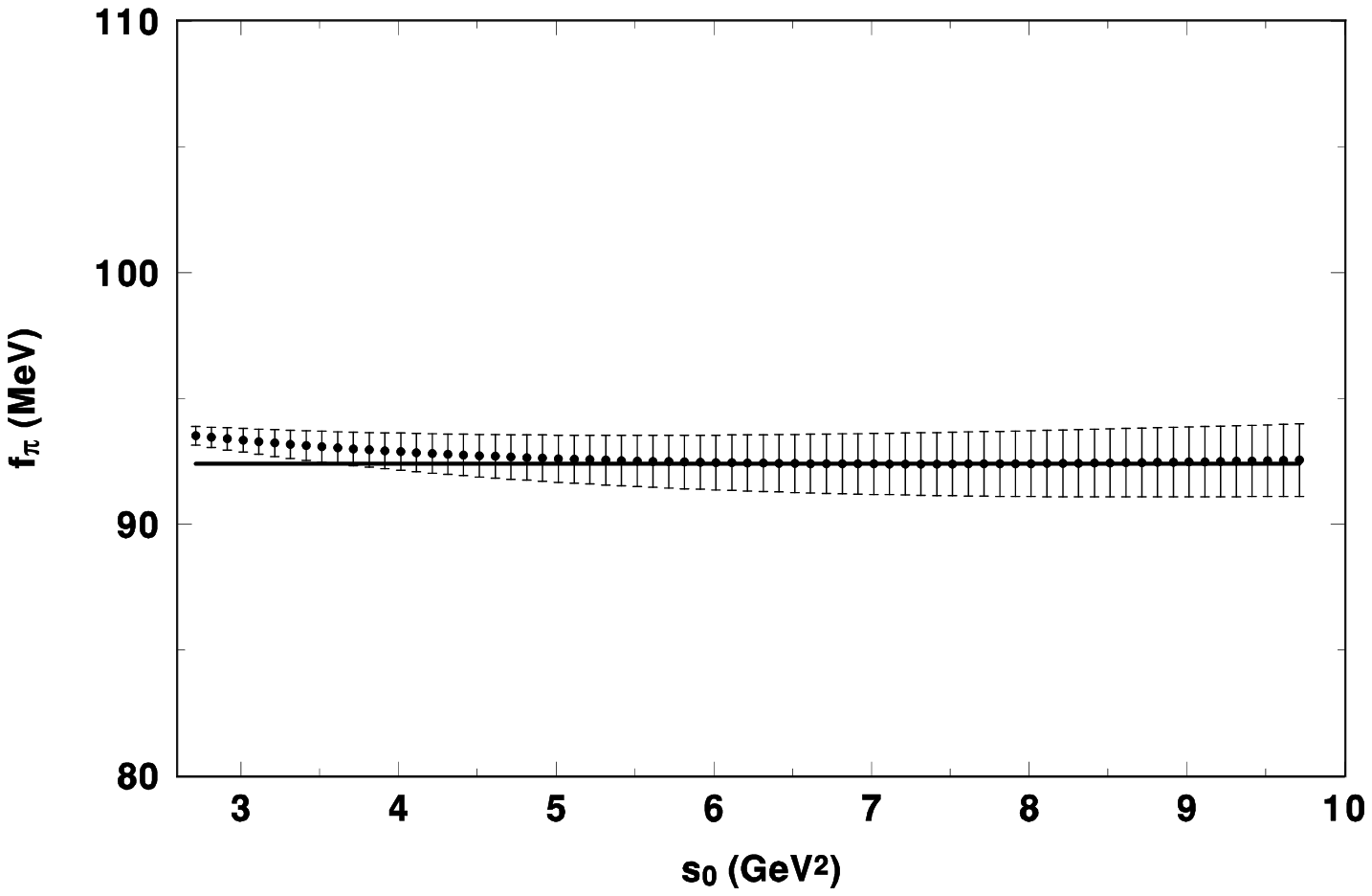}
  \caption{\footnotesize{Results for $f_\pi$ from the FESR in the axial-vector channel, Eq. (18), with $C_2\langle{\hat{O}}_2\rangle =0$, $C_4\langle{\hat{O}}_4\rangle = 0.05 \,\mbox{GeV}^4$, and using  CIPT  with $\Lambda_{QCD} = 365 \,\mbox{MeV}$ ($\alpha_s(M_\tau)= 0.335$). The straight line  is the experimental value of $f_\pi$, and the points with errors are the integrated data up to $s_1\simeq M_\tau$.}}
  \label{fig:figure6}
  \end{center}
\end{figure}
If the hadronic spectral function is known in some channel from experiment, e.g. from $\tau$-decay into hadrons, then $Im \,\Pi(s)|_{HAD} \equiv Im \,\Pi(s)|_{DATA}$, and Eq.(14) can be used to determine the values of the vacuum condensates. Subsequently, Eq.(14) can be used in a different channel for a different application. It is important to mention that the correlator $\Pi(q^2)$ is generally not a physical observable. However, this has no effect in FOPT as the unphysical quantities (polynomials) in the correlator do not contribute to the integrals. In the case of CIPT, though, this requires modified sum rules involving as many derivatives of the correlator as necessary to render it physical.\\
Next, let us consider  an application where the integration kernel $f(s)$ in Eq.(12) is of great importance \cite{FPI}. In the axial-vector channel, the FESR Eq.(14) with $f(s)=1$ and $N=0$ can be confronted with data from $\tau$-decay. 
The hadronic spectral function is then written as the sum of the pion pole and the resonance data known up to the kinematical end point $s_0 = M_\tau^2$. The moment $M_2(s_0)$ is known up to five-loop order in PQCD, so that the FESR can be used to confront the resonance data plus PQCD with e.g. $f_\pi$. As seen from Fig. 5 the agreement is rather poor, except possibly near the end point. At first sight, this may be interpreted as a signal for quark-hadron duality violations near the real axis, even at this high enough energy. In fact, it has been known for quite some time that e.g. the Weinberg (chiral) sum rules are not saturated by the $\tau$ decay data unless one introduces {\it pinched} integration kernels, e.g.  $f(s) = [1 - (s/s_0)]^{(N+1)}$ \cite{PINCH1}-\cite{PINCH2}. Unfortunately, the $\tau$-lepton is not massive enough to probe higher energy regions. In spite of this it is still possible to explore a wider  energy range by introducing as integration kernel a polynomial $f(s) \equiv P(s, s_0, s_1)$ tuned to eliminate the (unknown) hadronic contribution to the integral between $s_1$ and $s_0 \geq s_1$, where $s_1$ is at or near the end point of the data. It has been shown \cite{FPI} that the optimal degree of $P(s)$ is the simplest, i.e. the linear function
\begin{equation}
P(s,s_0,s_1)=1-\frac{2s}{s_{0}+s_{1}} \,,
\end{equation}
so that
\begin{equation}
 \mbox{constant} \times \int_{s_1}^{s_0} P(s,s_0,s_1) ds  = 0\,.
\end{equation}
In this case the complete FESR becomes a linear combination of a dimension-two and a dimension-four FESR, which from Eqs.(14) and (16)  is given by
\begin{eqnarray}
2 \, f_\pi^2 &=& - \int_{0}^{s_{1}} ds \, P(s) \, \frac{1}{\pi}\, Im \,\Pi(s)|_{DATA}
+ \frac{s_0}{4 \pi^2} \left[ M_2(s_0) - \frac{2 s_0}{s_0+s_1} M_4 (s_0) \right] \nonumber \\[.3cm]
&+& \frac{1}{4 \pi^2} \left[ C_2 \langle \hat{O}_2 \rangle +\frac{2}{s_0+s_1} C_4 \langle \hat{O}_4 \rangle \right] \, + \Delta(s_0)\,,
\end{eqnarray}
where the pion pole has been separated from the data, and the chiral limit is understood. The term $\Delta(s_0)$ is the error being made by assuming that the data is constant in the interval $s_1 - s_0$. It is possible to estimate this error which turns out to be two to three orders of magnitude smaller than  $2 f_\pi^2$ on the left hand side of Eq.(18) \cite{FPI}. 
As can be seen from Fig. 6 the FESR Eq.(18) shows an excellent consistency between QCD and the $\tau$ data in the axial-vector channel in a remarkably wide region $s_0 \simeq 4 \, -\,10\, \mbox{GeV}^2$. A similar consistency is also found in the vector channel, where QCD is now confronted with zero (there is no pole in this channel). This result shows either no evidence for quark-hadron duality violations in these channels, or if they are present it indicates that they are suppressed by the integration kernel  (some model dependent analyses claim the existence of duality violations \cite{PICH3}).\\ 

\section{Light quark masses}
Traditionally, the light quark masses have been determined using the correlator, Eq.(8), involving  the pseudoscalar currents $J(x) \equiv \partial_\mu A^\mu(x)|^i_j = [\overline{m}_i(\mu) + \overline{m}_j(\mu)]:\overline{q}_j(x) i \gamma_5 q_i(x):$, where $A_\mu(x)$ is the axial vector current of flavours $i$ and $j$, $\overline{m}_i(\mu)$ the quark mass  in the $\overline{MS}$ scheme, $\mu$ the renormalization scale and $q_i(x)$ are the quark fields. An issue of major concern in the past was the presence of logarithmic quark-mass singularities in these correlators. This problem  has been satisfactorily resolved some time ago in \cite{LMS1}-\cite{LMS2}.
These correlators are now known to five-loop order in PQCD \cite{CHET5}, and free of logarithmic quark mass singularities. The Wilson coefficients of  the leading power corrections, i.e. the gluon and the quark condensates, are also known up to two-loop level \cite{LMS1}-\cite{LMS2}. Higher dimensional condensates, as well as quark mass corrections of order ${\cal{O}}$$(m_i^4)$ (with respect to the one-loop term) and higher turn out to be negligible. From Cauchy's theorem, Eq.(12), the FESR  to determine the quark masses can be written as
\begin{eqnarray}
- \frac{1}{2\pi i}
\oint_{C(|s_0|)}
ds \;\psi_{5}^{QCD}(s)\; \Delta_5(s) &=& 2\; f_P^2 \; M_P^4\; \Delta_5(M_P^2)  \nonumber \\ [.3cm]
&+&
\int_{s_{th}}^{s_0}
ds \;\frac{1}{\pi} \;Im \;\psi_{5}(s)|_{RES}\;\Delta_5(s) \, , 
\end{eqnarray}
where $\Delta_5(s)$ is an (analytic) integration kernel to be introduced shortly,  the first term on the right hand side  is the pseudoscalar meson pole contribution ($P = \pi, K$), $s_{th}$ is the hadronic threshold, and $Im\, \psi_5(s)|_{RES}$ is the hadronic resonance spectral function. The radius of integration $s_0$ is assumed to be large enough for QCD to be valid on the circle. For later convenience this FESR can be rewritten as
\begin{equation}
\delta_5(s_0)|_{QCD} \,=\, \delta_5|_{POLE}\, + \, \delta_5(s_0)|_{RES}\;, 
\end{equation}
where the meaning of each term is self evident. Historically, the problem with the pseudoscalar correlator has been the lack of direct experimental information on the hadronic resonance spectral functions. Two radial excitations of the pion and of the kaon, with known masses and widths, have been observed in hadronic interactions \cite{PDG}. However, this information is hardly enough to reconstruct the full spectral function. In fact, inelasticity, non-resonant background and resonance interference are impossible to guess, leaving no choice but to model these functions. This introduces an unknown systematic uncertainty which has been present in all past QCD sum rule determinations of the light quark masses. Since the FESR Eq.(19) is valid for any analytic  $\Delta_5(s)$ one can choose this kernel in such a way as to suppress $\delta_5(s_0)|_{RES}$ as much as possible. An example of such a function is the second degree polynomial \cite{ss}-\cite{IJMPA}
\begin{equation}
\Delta_5(s)|_{RES} \,=\, 1 \, - a_0 \,s - a_1\, s^2\;, 
\end{equation}
where $a_0$ and $a_1$ are constants fixed by the requirement $\Delta_5(M_1^2) = \Delta_5(M_2^2) =0$, where $M_{1,2}$ are the masses of the first two radial excitations of the pion or kaon. This simple kernel suppresses enormously the resonance contribution, which becomes only a couple of a percent of the pole contribution, and well below the current uncertainty due to the strong coupling. This welcome feature is essentially independent of the model chosen to parametrize the resonances. A practical parametrization consists of two Breit-Wigner forms normalized at threshold according to chiral perturbation theory, as first proposed in \cite{CADCHPT1} for the pionic channel, and in \cite{CADCHPT2} for the kaonic channel.
Detailed results for $\delta_5(s_0)|_{QCD}$, to five-loop order in PQCD and up to dimension $d=4$ in the OPE, after integrating in FOPT may be found in \cite{ms}.  
In the case of CIPT the FESR must be written in terms of the second derivative of the current correlator. This is in order to eliminate the unphysical first degree polynomial present in $\psi_5(s)$, which unlike the case of FOPT would otherwise contribute to the FESR which then becomes
\begin{eqnarray}
- \frac{1}{2\pi i}
\oint_{C(|s_0|)}
&ds& \psi_{5}^{'' QCD}(s)\,[F(s) - F(s_0)] = 2\; f_P^2 \; M_P^4\; \Delta_5(M_P^2)  \nonumber \\ [.3cm]
&+&
\frac{1}{\pi} \; \int_{s_{th}}^{s_0}
ds \; Im \;\psi_{5}(s)|_{RES}\;\Delta_5(s) \, , 
\end{eqnarray}
where
\begin{equation}
F(s) = - s \left(s_0 - a_0\,\frac{s_0^2}{2} - a_1\, \frac{s_0^3}{3} \right) + \frac{s^2}{2} - a_0\, \frac{s^3}{6} - a_1\, \frac{s^4}{12} \;,
\end{equation}
and
\begin{equation}
F(s_0) = - \frac{s_0^2}{2} +  a_0\, \frac{s_0^3}{3} + a_1\, \frac{s_0^4}{4} \;.
\end{equation}
\begin{figure}[ht]
\begin{center}
  \includegraphics[height=.35\textheight]{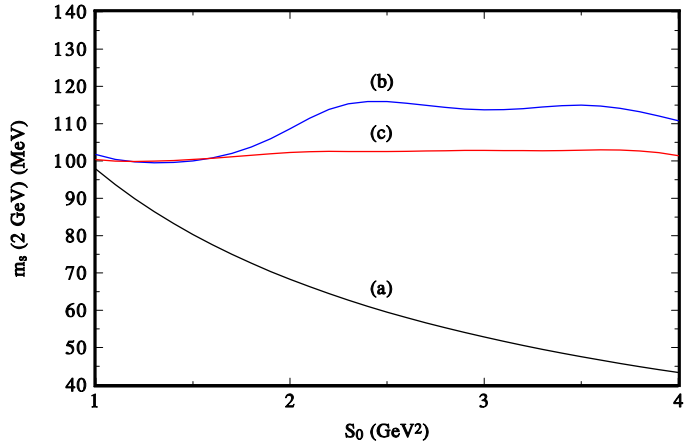}
  \caption{\footnotesize{The strange quark mass $\overline{m}_s (2\; \mbox{GeV})$ in the $\overline{MS}$ scheme taking into account only the kaon pole with $\Delta_5(s) = 1$ (curve(a)), and the two Breit-Wigner resonance spectral function with a threshold constraint from CHPT \cite{CADCHPT2}, with $\Delta_5(s) = 1$ (curve(b)), and $\Delta_5(s)$ as in Eq.(21) (curve(c)). A systematic uncertainty of some $20 \,\%$  due to the resonance sector is dramatically unveiled.}}
  \label{fig:figure7}
  \end{center}
\end{figure} 
The RG improvement is used before integration, so that all logarithmic terms vanish. The running coupling as well as the running quark masses are no longer frozen as in FOPT, but must be integrated. This can be done by solving numerically the respective RG equations at each point on the integration circle in the complex s-plane. Detailed expressions are given in \cite{ms}-\cite{mq}.\\
The parameters of the integration kernel, Eq.(21), are $a_0 = 0.897 \;\mbox{GeV}^{-2}$, and $a_1 = -\, 0.1806 \;\mbox{GeV}^{-4}$ for the pionic channel, and $a_0 = 0.768 \;\mbox{GeV}^{-2}$, and $a_1 = -\, 0.140 \;\mbox{GeV}^{-4}$ for the kaonic channel. These values correspond to the radial excitations $\pi(1300)$, $\pi(1800)$, $K(1460)$ and $K(1830)$. The pion and kaon decay constants are \cite{PDG} $f_\pi = 92.21 \,\pm 0.14 \;\mbox{MeV}$, and $f_K = (1.22 \pm 0.01) f_\pi$. In the QCD sector it is best to use the value of the strong coupling determined at the scale of the $\tau$-mass, as this is close to the scale in current use for the light quark masses, i.e. $\mu= 2 \;\mbox{GeV}$. The extraction of $\alpha_s(M_\tau)$ from the $R_\tau$ ratio  involves an integral with a natural kinematical integration kernel that eliminates the contribution of the $d=4$ term in the OPE. This welcome feature improves the accuracy of the determination, and it makes little sense to introduce additional spurious integration kernels which would artificially recover this $d=4$ contribution. The different values obtained from $\tau$ decay using CIPT are all in agreement with each other, i.e. $\alpha_s(M_\tau) = 0.338 \pm 0.012$ \cite{PICH2}, $\alpha_s(M_\tau) = 0.341 \pm 0.008$ \cite{CVETIC}, $\alpha_s(M_\tau) = 0.344 \pm 0.009$ \cite{DAVIER}, and $\alpha_s(M_\tau) = 0.332 \pm 0.016$ \cite{BAIKOV}. These determinations are model independent and
extremely transparent, with $\alpha_s$ obtained essentially by confronting PQCD with the single experimental number $R_\tau$. The method of FOPT is known to give rise to a pathological non-convergent perturbative series in this application \cite{PICH1}, so it will not be considered here. The $d=4$ gluon condensate has been extracted from $\tau$ decays \cite{C2b}, but one can conservatively consider the wide range$<\alpha_s G^2> = 0.01 - 0.12 \;\mbox{GeV}^4$. The impact of the light quark condensate is at the level of $1 \%$ in the quark masses. A $\pm \;30 \%$ uncertainty in the resonance contribution $\delta_5(s_0)|_{RES}$ in Eq.(20) translates into a safe $1 \%$ change in the quark masses. Finally, it has been assumed that the unknown six-loop PQCD contribution is equal to the five-loop result, an extreme but very conservative estimate of higher orders in PQCD.\\ 
\begin{table}
\small
\begin{tabular}{ccccccccc}
\hline \\
\noalign{\smallskip}
Source & $\overline{m}_u$ & $\overline{m}_d$& $\overline{m}_s$& $\overline{m}_{ud}$ & $\overline{m}_u/\overline{m}_d$ & $\overline{m}_s/\overline{m}_{ud}$ & $R$ & $Q$ \\
\hline \\
\noalign{\smallskip}
QCDSR \cite{mq}  &  $2.6 \pm 0.3$ & $5.6 \pm 0.4$ &  - & $4.1 \pm 0.3$   &  $0.46 \pm 0.06$ & -  &- &  -\\
\hline \\
\noalign{\smallskip}
QCDSR \cite{ms}  &  - & - & $ 102 \pm 8$  & -  & - & $24.9 \pm 2.7$  &$33 \pm 6$ & $21 \pm 3$ \\
\hline \\
\noalign{\smallskip}
FLAG \cite{CHPT2}  &  $2.2 \pm 0.3$ &  $4.6 \pm 0.6$  & $ 95 \pm 10$ &  $3.4 \pm 0.4$   & $0.47 \pm 0.04$  & $27.8 \pm 1.0$  & $37.2 \pm 4.1$ & $23.1 \pm 1.5$ \\
\noalign{\smallskip}
\hline
\end{tabular}
\caption{\footnotesize{The running quark masses in the $\overline{MS}$ scheme at a scale $\mu = 2 \; \mbox{GeV}$ in units of MeV from QCD sum rules (QCDSR) (first two rows), and from the FLAG lattice QCD analysis \cite{CHPT2}. The ratios $\overline{m}_u/\overline{m}_d$ and  $\overline{m}_s/\overline{m}_{ud}$ are an input in the QCDSR (see text). The ratios $R$ and $Q$ are defined on the l.h.s. of  Eqs.(2) and (3). The QCDSR results for the up- and down- quark masses and ratios differ slightly from \cite{mq}-\cite{IJMPA} due to the input value of $\overline{m}_u/\overline{m}_d$ used here.}}
\end{table}
\normalsize
Beginning with the strange quark mass,  Fig. 1 shows the results for $\overline{m}_s (2\; \mbox{GeV})|_{\overline{MS}}$ with no integration kernel, $\Delta_5(s) = 1$, and taking into account only the kaon pole, curve (a), and the kaon pole plus a two Breit-Wigner resonance model with a threshold constraint from CHPT \cite{CADCHPT2}, curve (b) (a misprint in the formula for the spectral function in \cite{CADCHPT2} has been corrected in  \cite{CADCHPT3}). These curves are for the central value of $\alpha_s(M_\tau)$ whose uncertainties will be considered afterwards.
The latter result is reasonably stable in the wide region $s_0 = 2 - 4 \; \mbox{GeV}^2$, so that it could lead us to conclude that $\overline{m}_s (2\; \mbox{GeV})|_{\overline{MS}}  \simeq 100 - 120 \; \mbox{MeV}$, albeit with a yet unknown systematic uncertainty arising from the resonance sector.
Introducing the kernel, Eq.(21), leads to curve (c) and to a dramatic unveiling of this systematic uncertainty. In fact, the {\it real} value of the quark mass is $\overline{m}_s (2\; \mbox{GeV})|_{\overline{MS}} = 102 \pm 8 \; \mbox{MeV}$, or some $20 \%$ below the former result (this error now includes the uncertainty in $\alpha_s$). In addition, and as a bonus the systematic uncertainty-free result is remarkably stable in the  unusually wide region $s_0  \simeq 1 - 4 \; \mbox{GeV}^2$ (typical stability regions are only half as wide).\\ It must be recalled that the pseudoscalar correlator involves the overal factor $(m_s + m_{ud})^2$. Hence, in order to determine $m_s$ an input value for the ratio $m_s/m_{ud}$ is needed in the result from the sum rule, which is
\begin{equation}
\overline{m}_s (2\; \mbox{GeV})|_{\overline{MS}} = \frac{105.5 \pm 8.2 \; \mbox{MeV}}{1 + m_{ud}/m_s} \;.
\end{equation}
Using the wide range $m_s/m_{ud} = 24 - 29$ leads to $\overline{m}_s (2\; \mbox{GeV})|_{\overline{MS}} = 102 \pm 8 \; \mbox{MeV}$. In this case the impact of the uncertainty in the quark mass ratio is  small. However, in the case of the up- and down-quark masses  the corresponding ratio $m_u/m_d$ plays a more important role in the result from the sum rule, which is
\begin{equation}
\overline{m}_d (2\; \mbox{GeV})|_{\overline{MS}} = \frac{8.2 \pm 0.6 \; \mbox{MeV}}{1 + m_u/m_d} \;.
\end{equation}
The input used  in Eq.(26) for the ratio $m_u/m_d = 0.47 \pm 0.04$  is from the overall lattice QCD analysis of the FLAG group \cite{CHPT2}. Once $m_d$ is determined from Eq.(26),  $m_u$ follows. Using these results for the individual masses one obtains the ratios $m_u/m_d$ and $m_s/m_{ud}$ shown in Table 1.
The quark masses $\overline{m}_u (2\; \mbox{GeV})|_{\overline{MS}}$ and $\overline{m}_d (2\; \mbox{GeV})|_{\overline{MS}}$
also exhibit a remarkably wide stability region $s_0 \simeq 1 - 4 \; \mbox{GeV}^2$ \cite{mq}-\cite{IJMPA}.
In Table 1  one  finds a summary of the results for the light quark masses, and the ratios $R$ and $Q$ defined on the left hand side of Eqs. (2) and (5), together with the results of the Flag group \cite{CHPT2}. The values of the up- and down-quark masses and their ratios in Table 1 are slightly different from those in \cite{mq}-\cite{IJMPA} due to the input value for the ratio $m_u/m_d$ (in
\cite{mq}-\cite{IJMPA} the value $m_u/m_d = 0.55$ was used).
The various sources of errors in the quark masses discussed earlier combine into the final values given in Table 1. Having all but eliminated the systematic uncertainty from the hadronic resonance sector, the main source of error is now due to the strong coupling. Improved accuracy in the determination of $\alpha_s$ would then allow for a reduction of the uncertainties in the light quark masses.
\section{Heavy quark masses}
Determinations of the charm- and bottom-quark masses are not affected by a lack of data, as there is plenty of experimental information from $e^+ e^-$ annihilation into hadrons \cite{PDG}, except for a gap in the region
$25 \;\mbox{GeV}^2 \lesssim s \lesssim 50\; \mbox{GeV}^2$ . On the theoretical side there has been very good progress on PQCD  up to four-loop level \cite{QCD1}-\cite{QCD14}. The leading power correction in the OPE is due to the gluon condensate with its Wilson coefficient known at the two-loop level \cite{BROAD}.
The correlator, Eq.(8), involves the vector current $J(x)\equiv V_\mu(x) = \bar{Q}(x) \gamma_\mu Q(x)$, where $Q(x)$ is the charm- or bottom-quark field. The experimental data is in the form of the $R_Q$-ratio for charm (bottom) production, which determines the hadronic spectral function. Modern determinations of the heavy-quark masses have been based on inverse moment (Hilbert-type) QCD sum rules, e.g. Eq.(12) with $f(s) = 1/s^n$. These sum rules require QCD knowledge of the vector correlator in the low energy region, around the open charm (bottom) threshold, as well as in the high energy region. A recent update \cite{KUHN10} of earlier determinations \cite{QCD1}-\cite{QCD3}, \cite{QCD5}-\cite{QCD7} reports a charm-quark mass in the $\overline{MS}$ scheme accurate to $1 \%$, and half this uncertainty for the bottom-quark mass. However, the analysis of \cite{QCD14} claims an error a factor two larger for the charm-quark mass. It appears that the discrepancy arises from the treatment of PQCD. In fact, in \cite{QCD14} two different renormalization scales were used, one for the strong coupling and another one for the quark mass. This unconventional choice results in a much larger error in the charm-quark mass obtained from  inverse (Hilbert) moment QCD sum rules. It does not affect, though, sum rules involving positive powers of $s$. In any case, the philosophy in current use is to choose the result from the method leading to the smallest uncertainty. \\ 
Beginning with the charm-quark mass, an alternative procedure was proposed some years  ago based  only on the high energy expansion of the heavy-quark vector correlator \cite{KS1}-\cite{KS2}. This method was followed recently \cite{SB1}, but with updated PQCD information and the inclusion of  integration kernels in the FESR, Eq.(12), tuned to enhance/suppress contributions from data in certain regions. The first such kernel is  the so-called {\it pinched} kernel \cite{PINCH1}-\cite{PINCH2}
\begin{equation}
f(s) = 1 - \frac{s}{s_0} \;,
\end{equation}
which is supposed to suppress potential duality violations close to the real s-axis in the complex s-plane. In connection with the charm-quark mass application, this kernel enhances the contribution from the first two narrow resonances, $J/\psi$ and $\psi(2S)$, and reduces the weight of the broad resonance region, particularly near the onset of the continuum. The latter feature is better achieved with the alternative kernel \cite{SB2}
\begin{equation}
f(s) = 1 - \left(\frac{s_0}{s}\right)^2 \;,
\end{equation}
which produces an obvious larger enhancement of the narrow resonances, and a larger quenching of the broad resonance region. This kernel has been used together with both the high and the low energy expansion of the vector correlator in \cite{SB2}. Since there are no data in the region  $25 \;\mbox{GeV}^2 \lesssim s \lesssim 50\; \mbox{GeV}^2$, while the data for $s\gtrsim 50\;\mbox{GeV}^2$ agrees with PQCD, it is useful to introduce a kernel that will allow a suppression in the former region, e.g. 
\begin{equation}
f(s) = \mathcal{P}_n [x(s)] \;,
\end{equation}
where
\begin{equation}
x(s) = \frac{2 s - (s_0 + s_1)}{s_0 - s_1}\;,
\end{equation}
with $s_0 > s_1$, and $\mathcal{P}_n(x)$ are the standard Legendre polynomials, i.e. $\mathcal{P}_1(x) = x$, $\mathcal{P}_2(x) =(5 x^3 - 3 x)/2$, etc., which satisfy the constraint
\begin{equation}
\int_{s_1}^{s_0} s^k \;\mathcal{P}_n[x(s)]\;ds =0 \;,
\end{equation}
where $s_1 \simeq 24 \; \mbox{GeV}^2$, and $s_0$ varies in the region where there is no data. 
These Legendre-type kernels provide extra weight to the well known resonance region on account of their rapid growth for $s < s_1$. The resulting charm-quark masses are essentially insensitive to the order of these polynomials, with $n=3-6$ giving answers differing at the $0.1 \%$ level.\\
\begin{table}
\begin{center}
\small
\begin{tabular}{cccccccc}
\hline \\
\multicolumn{7}{r}{Uncertainties (in MeV)} \\
\cline{3-8}
\noalign{\smallskip}
 Kernel & $\overline{m}_c(3\,\mbox{GeV})$ & EXP & $\Delta \alpha_s$  & $\Delta \mu$ & NP & $s_0$ &  Total               \\
\hline\\
\noalign{\smallskip}
$1 - s/s_0$ & 983 \quad & \quad 9 \quad & \quad 1 \quad & \quad - \quad & \quad 1 \quad & \quad 16 &\quad 25\\
\hline\\
\noalign{\smallskip}
$\mathcal{P}_{5}[x(s)]$ & 1007 \quad & \quad 22 \quad & \quad 1 \quad & \quad 8 \quad & \quad 2 \quad & \quad $< 1$  & \quad 23 \\
\hline\\
\noalign{\smallskip}
$s^{-2}$  &  995 \quad & \quad 9 \quad &\quad  3 \quad &\quad  1 \quad  &\quad  1 \quad &\quad 14 &\quad 17\\
\hline\\
\noalign{\smallskip}
$1-(s_0/s)^2$  &  987 \quad & \quad 7 \quad &\quad  4 \quad &\quad  1 \quad  &\quad  1 \quad &\quad 4 &\quad 9\\
\hline\\
\end{tabular}
\caption{\footnotesize{Results for $\overline{m}_c(3\,\mbox{GeV})$ in the $\overline{MS}$ scheme and in MeV from different integration kernels in the FESR. Legendre polynomial kernels for $n=3-6$ give basically the same result as for $n=5$. The various uncertainties are due to the data (EXP), the value of $\alpha_s$ ($\Delta \alpha_s$), changes of $\pm 35 \%$ in the renormalization scale around $\mu= 3 \; \mbox{GeV}$ ($\Delta \mu$), the value of the gluon condensate (NP), and due to variations of $s_0$ ($s_0$) in a very wide range (see text). The first four errors are added in quadrature, but the last one is conservatively added linearly. The first two rows are from \cite{SB1} and the last two from \cite{SB2}. The total error in the first row includes an estimate of the error due to PQCD truncation at the four-loop level ($ 2 \; \mbox{MeV}$). CIPT was used throughout.}}
\end{center}
\end{table}
The results for $\overline{m}_c(3\,\mbox{GeV})$ in the $\overline{MS}$ scheme using various integration kernels are listed in Table 2, together with the various uncertainties. The merits of each kernel may be judged by its ability to minimize these uncertainties, in particular those that might be most affected by systematic errors, such as  e.g. the experimental data. The kernel $f(s) = 1 - (s_0/s)^2$ appears to be optimal as it produces the smallest uncertainty due to the data, and is very stable against changes in $s_0$. Some recent determinations of the charm-quark mass  are based on Hilbert moments with no  $s_0$-dependent kernel, corresponding to the third row in Table 2. While  there is no explicit $s_0$-dependence in Hilbert moments (the integrals extend to infinity), there is definitely a residual dependence when choosing the threshold for the onset of PQCD. From a FESR perspective, the major drawback of the kernel $1/s^2$ is clearly the poor stability against changes in $s_0$. In this regard, the Legendre polynomial kernels would be optimal, except that they have the largest uncertainty due to the data. An important remark is in order concerning the uncertainty due to changes in $s_0$. From current data it is not totally clear where does PQCD start. This problem not only affects FESR, with their explicitly obvious $s_0$-dependence, but also Hilbert moments with an implicit $s_0$-dependence, as there is no data all the way up to infinity. The range of variation in $s_0$ and its contribution to the uncertainty in the quark mass, as appearing in Table 2, is as follows. For rows 1, 3 and 4 
$s_0 \simeq 15 - 23 \; \mbox{GeV}^2$, and for the Legendre polynomial kernels, row 2, it is $s_0 \simeq 100 - 200 \; \mbox{GeV}^2$.
Two of the most recent results for $\overline{m}_c(3\,\mbox{GeV})$ \cite{KUHN10}, \cite{QCD14}, together with the weighted FESR value \cite{SB2} (last row in Table 2) are 
\begin{eqnarray}
\overline{m}_c(3\,\mbox{GeV}) \;  = \;\Bigg\{ 
\begin{array}{lcl}
986  \; \pm 13 \;\mbox{MeV} \; \cite{KUHN10}\\
998  \; \pm 29 \;\mbox{MeV} \; \cite{QCD14} \\
987  \; \pm \;\; 9 \;\mbox{MeV} \; \cite{SB2}\;, 
\end{array}
\end{eqnarray}
in very good agreement with each other, except for the errors. The small uncertainty from \cite{SB2} is due in part to improved quenching of the data in the broad resonance region, but mostly due to a strong reduction in the sensitivity to $s_0$, i.e.  the onset of PQCD.  For comparison, a recent lattice QCD determination gives \cite{LATT1}
\begin{equation}
\overline{m}_c(3\,\mbox{GeV}) \;  = \; 986 \; \pm\; 6 \;\mbox{MeV}\;.
\end{equation}
Turning to the bottom-quark mass, the most recent determination from Hilbert moment QCD sum rules  \cite{KUHN10} is
\begin{equation}
\overline{m}_b(10\,\mbox{GeV}) \;  = \; 3610 \; \pm\; 16 \;\mbox{MeV}\;.
\end{equation}
The error above is due to uncertainties in the data, the strong coupling and the renormalization scale. It does not include, though, an uncertainty due to the onset of PQCD. There is an appreciable difference if PQCD were to start at the end point of the (BABAR) data ($\sqrt{s_0} \simeq 11.2\; \mbox{GeV}$), or rather at higher energies ($\sqrt{s_0} \simeq 13.0 \;\mbox{GeV}$).
If one were to adopt the procedure of \cite{SB1}, i.e. add in quadrature the first three errors, but only linearly the one due to $s_0$, then using the information in \cite{KUHN10} the result would change to $\overline{m}_b(10\;\mbox{GeV}) \;  = \; 3600 \; \pm\; 47 \;\mbox{MeV}$. Presumably, the error would be larger if a broader region in $s_0$ were to be considered. One of the virtues of weighted sum rules is to reduce this uncertainty, as may appreciated from Table 2 for the case of the charm-quark mass. For the bottom-quark mass work on weighted FESR is in progress \cite{bquark}. 
In any case, it should be kept in mind that by choosing different kernels it is possible to generate a large number of predictions for the quark masses with a variety of different errors. According to current philosophy  one chooses the determination having the smallest error.

\section{Conclusions}
After a short review of quark mass ratios the method of QCD sum rules was discussed,  in connection with determinations of individual values of the quark masses. The historical (unknown) hadronic systematic uncertainty affecting light quark mass determinations was highlighted. Details of the recent breakthrough in eliminating this uncertainty were provided. Future improvement in accuracy is now possible, and depends essentially on more accurate determinations of the strong coupling, now the main source of error. The new values of the light quark masses and their ratios, free from this systematic uncertainty, agree well with lattice QCD results. In the heavy quark sector recent more accurate determinations of the charm- and bottom-quark masses were reported. While these values are all in agreement, there is some disagreement on the size of the errors. The use of suitable multipurpose integration kernels in FESR allows to tune the weight of the various contributions to the quark masses. This in turn allows to minimize the error due to the data, as well as to the uncertainty in the onset of PQCD. The latter uncertainty impacts FESR as well as Hilbert moment sum rules, as there is no data all the way up to infinity. If no kernel, other than inverse powers of $s$ is used then this uncertainty would be much larger than normally reported, as may be appreciated from Table 2. However, according to  current philosophy one chooses the determination having the smallest error.
Marginal improvement of the current total error in this framework should be possible with improved accuracy in the data and in the strong coupling.
 
\section{Acknowledgements}
The author wishes to thank his collaborators  in the various projects on quark masses reported here: S. Bodenstein, J. Bordes, N. Nasrallah, J. Pe\~{n}arrocha, R. R\"{o}ntsch and  K. Schilcher. Enlightening correspondence with H. Leutwyler and P. Minkowski is greatly appreciated. This work was supported in part by NRF (South Africa).
 


\begin{thebibliography}{9}

\bibitem{QCDSR}  P. Colangelo and  A. Khodjamirian, in: "At the Frontier of Particle Physics/ Handbook of QCD"', M. A. Shifman, ed. (World Scientific, Singapore 2001), Vol. 3, 1495.
\bibitem{PICH1} A. Pich, Acta Phys. Polon. Supp. {\bf 3}, 165 (2010).
\bibitem{PICH2} A. Pich, arXiv:1101.2107.
\bibitem{CA1} S. Glashow and S. Weinberg, Phys. Rev. Lett. {\bf 20}, 224 (1968).
\bibitem{CA2} M. Gell-Mann, R. Oakes and B. Renner, Phys. Rev. {\bf 175}, 2195 (1968).
\bibitem{HP} H. Pagels, Phys. Rep.  C {\bf 16}, 219 (1975).
\bibitem{HLR} J. Gasser and H. Leutwyler, Phys. Rep. C {\bf 87}, 77 (1982).
\bibitem{CAD1}  C. A. Dominguez and A. Zepeda, Phys. Rev. D {\bf 18}, 884 (1978).
\bibitem{CAD2} C. A. Dominguez, Phys. Lett. B {\bf  86}, 171 (1979).
\bibitem{MZ} P. Minkowski and A. Zepeda, Nucl. Phys. B {\bf 164}, 25 (1980).
\bibitem{CHPT1} H. Leutwyler, PoS CD {\bf 09}, 005 (2009).
\bibitem{CHPT2}G. Colangelo {\it et al.}, arXiv:1011.4408.
\bibitem{SW} S. Weinberg, Trans. New York Acad. Sci. {\bf 38}, 185 (1977).
\bibitem{GC} G. Colangelo, S. Lanz, and E. Passemar, PoS  CD  {\bf 09}, 047 (2009).
\bibitem{eta} H. Leutwyler, Nucl. Phys. B Proc. Suppl. {\bf 64}, 223 (1998).
\bibitem{C2a} C. A. Dominguez and K. Schilcher, Phys. Rev. D \textbf{61}, 114020 (2000).
\bibitem{C2b} C. A. Dominguez and K. Schilcher, J. High Energy Phys. \textbf{0701}, 093 (2007).
\bibitem{GMOR}J. Bordes, C. A. Dominguez, P. Moodley, J. Pe\~{n}arrocha, and K. Schilcher, J. High Energy Phys. \textbf {1005}, 064 (2010).
\bibitem{Shankar} R. Shankar, Phys. Rev. D {\bf 15}, 755 (1977).
\bibitem{FPI} C. A. Dominguez, N. F. Nasrallah, and K. Schilcher, Phys. Rev. D  \textbf{80}, 054014 (2009).
\bibitem{PINCH1} K. Maltman, Phys. Lett. B {\bf 440}, 367 (1998).
\bibitem{PINCH1b} C. A. Dominguez and K. Schilcher, Physics Letters B  \textbf{448}, 93 (1999).
\bibitem{PINCH2} C. A. Dominguez and K. Schilcher, Physics Letters B  \textbf{581}, 193 (2004). 
\bibitem{PICH3} M. Gonzalez-Alonso, A. Pich and J. Prades, Phys. Rev. D {\bf 82}, 014019 (2010).
\bibitem{LMS1} K. G. Chetyrkin, C. A. Dominguez, D. Pirjol and K. Schilcher, Phys. Rev. D {\bf 51}, 5090 (1995).
\bibitem{LMS2}M. Jamin and M. M\"{u}nz, Z. Physik, C {\bf 66}, 633 (1995).
\bibitem{CHET5} P. A. Baikov, K. G. Chetyrkin and J. H. K\"{u}hn, Phys. Rev. Lett. {\bf 96}, 012003 (2006).
\bibitem{PDG} K. Nakamura et al., Particle Data Group, J. Phys. G {\bf 37}, 075021 (2010).
\bibitem{ss}C. A. Dominguez, N. F. Nasrallah, and K. Schilcher, J. High Energy Phys. \textbf {0802}, 072 (2008).
\bibitem{ms} C. A. Dominguez, N. F. Nasrallah, R. R\"{o}ntsch and K. Schilcher, J. High Energy Phys. \textbf{0805}, 020 (2008).
\bibitem{mq}C. A. Dominguez, N. F. Nasrallah, R. R\"{o}ntsch and and K. Schilcher, Phys. Rev. D {\bf 79}, 014009 (2009).
\bibitem{IJMPA} C. A. Dominguez, Int. J. Mod. Phys. A {\bf 29}, 5223 (2010).
\bibitem{CADCHPT1} C. A. Dominguez, Z. Phys. C {\bf 26}, 269 (1984).
\bibitem{CADCHPT2} C. A. Dominguez, L. Pirovano and K. Schilcher, Phys. Lett. B {\bf 425}, 193 (1998).
\bibitem{CADCHPT3}C. A. Dominguez, A. Ramlakan and K. Schilcher, Phys. Lett. B {\bf 511}, 59 (1998).
\bibitem{CVETIC}G. Cvetic {\it et al.}, Phys. Rev. D {\bf 82}, 093007 (2010).
\bibitem{DAVIER} M. Davier {\it et al.}, Eur. Phys. J. C {\bf 56}, 305 (2006).
\bibitem{BAIKOV} P. A. Baikov, K. G. Chetyrkin and J. H. K\"{u}hn, Phys. Rev. Lett. {\bf 101}, 012002 (2008).
\bibitem{QCD1}K. G. Chetyrkin, R. Harlander, J. H. K\"{u}hn, and M. Steinhauser,  Nucl. Phys. B {\bf 503}, 339 (1997).
\bibitem{QCD2}P. A. Baikov, K. G. Chetyrkin, and J. H. K\"{u}hn, Nucl. Phys. B (Proc. Suppl.) {\bf 189}, 49 (2009).
\bibitem{QCD3}K. G. Chetyrkin, R. Harlander, J. H. K\"{u}hn, Nucl. Phys. B {\bf 586}, 56 (2000).
\bibitem{QCD4}Y. Kiyo, A. Maier, P. Maierh\"{o}fer, and P. Marquard, Nucl. Phys. B {\bf 823}, 269 09). 
\bibitem{QCD5} P. A. Baikov, K. G. Chetyrkin, and J. H. K\"{u}hn, Phys. Rev. Lett. {\bf 101}, 012002 (2008).
\bibitem{QCD6} P. A. Baikov, K. G. Chetyrkin, and J. H. K\"{u}hn, Nucl. Phys. B (Proc. Suppl.) {\bf 135}, 243 (2004).
\bibitem{QCD7} K. G. Chetyrkin, J. H. K\"{u}hn, and M. Steinhauser, Phys. Lett. B {\bf 371}, 93 (1996).
\bibitem{QCD7b} K. G. Chetyrkin, J. H. K\"{u}hn, and M. Steinhauser, Nucl. Phys. B {\bf 482}, 213 (1996)
\bibitem{QCD7c} K. G. Chetyrkin, J. H. K\"{u}hn, and M. Steinhauser, Nucl. Phys. B {\bf 505}, 40 (1997).
\bibitem{QCD8} R. Boughezal, M. Czakon, and T. Schutzmeier, Phys. Rev. D {\bf 74}, 074006 (2006); Nucl. Phys. B (Proc. Suppl.) {\bf 160}, 164 (2006).
\bibitem{QCD9}A. Maier, P. Maier\"{o}fer, and P. Marquard, Nucl. Phys. B {\bf 797}, 218 (2008)
\bibitem{QCD9a} A. Maier, P. Maier\"{o}fer, and P. Marquard, Phys. Lett. B {\bf 669}, 88 (2008).
\bibitem{QCD10}K. G. Chetyrkin, J. H. K\"{u}hn, and C. Sturm, Eur. Phys. J. C {\bf 48}, 107 (2006).
\bibitem{QCD11} A. Maier, P. Maierh\"{o}fer, P. Marquard, and A. V. Smirnov, Nucl. Phys. B {\bf 824}, 1 (2010). 
\bibitem{QCD12}A. H. Hoang, V. Mateu, and S. Mohammad Zebarjad, Nucl. Phys. B {\bf 813}, 349 (2009).
\bibitem{QCD13}J. Hoff and M. Steinhauser, arXiv: 1103.1481.
\bibitem{QCD14} B. Dehnadi, A. H. Hoang, V. Mateu and S. Mohammad Zebarjad, arXiv: 1102.2264.
\bibitem{BROAD}D. J. Broadhurst {\it et al.} Phys. Lett. B {\bf 329}, 103 (1994).
\bibitem{KUHN10} K. G. Chetyrkin, {\it et al.}, arXiv: 1010.6157. 
\bibitem{KS1}J. Pe\~{n}arrocha, and K. Schilcher, Phys. Lett. B {\bf 515}, 291 (2001).
\bibitem{KS2} J. Bordes, J. Pe\~{n}arrocha, and K. Schilcher, ibid. B {\bf 562}, 81 (2003).
\bibitem{SB1} S. Bodenstein, J. Bordes, C. A. Dominguez, J. Pe\~{n}arrocha, and K. Schilcher, Phys. Rev. D {\bf 82}, 114013 (2010).
\bibitem{SB2} S. Bodenstein, J. Bordes, C. A. Dominguez, J. Pe\~{n}arrocha, and K. Schilcher, arXiv: 1102.3835, and Phys. Rev. D (in press).
\bibitem{LATT1} C. McNeile, C. T. H. Davies, E. Follana,. K. Hornbostel and G. P. Lepage, Phys. Rev. D {\bf 82}, 034512 (2010).
\bibitem{bquark}S. Bodenstein, J. Bordes, C. A. Dominguez, J. Pe\~{n}arrocha, and K. Schilcher, work in progress.
\end{thebibliography}
\end{document}